\documentclass{ptapap}

\author{Grzegorz Gajda}[CAMK]
\author{Ewa L. {\L}okas}[CAMK]
\author{E. Athanassoula}[LAM]
\affil[CAMK]{Nicolaus Copernicus Astronomical Center, Bartycka 18, 00--716 Warszawa, Poland}
\affil[LAM]{Laboratoire d'Astrophysique de Marseille, UMR6110, CNRS/Universit\'e de Provence,
38 rue Fr\'ed\'eric Joliot Curie, F-13388 Marseille C\'edex 13, France}

\title{Spectral analysis of stellar orbits\\ in a tidally induced bar}

\def \ptaJournalDate{to be published}
\def \ptaJournalVolume{vol.~2}
\def \ptaArticleURL{}
\def \ptaJournalName{PTA Proceedings}
\def \ArticleFirstPage{1}
\begin{document}

\maketitle

\begin{abstract}
Using numerical analysis of fundamental frequencies we study the orbital structure of a tidally induced bar
formed in a simulated dwarf galaxy orbiting a Milky Way-like host. We find that only about $10\%$ of stars
have frequencies compatible with x$_1$ orbits, the classical periodic orbits in a barred potential.
The rest of the stars follows box orbits parallel to the bar, with varying degree of elongation.
\end{abstract}

\section{Introduction}

Bars are common features of disk galaxies in the present Universe.
The estimated fraction of barred galaxies depends on the criteria used and ranges from $1/4$ to $2/3$
\citep[and references therein]{cheung13}.
Some of the dwarf galaxies in the Local Group possess a bar as well (e.g. the Large Magellanic Cloud) or are
supposed to have one \citep[e.g. the Sagittarius dwarf, see][]{lokas10}. Such bars may result from the tidal
interaction of initially disky dwarf irregular galaxies with a bigger host as an intermediate stage on the
way to their transformation into dwarf spheroidal galaxies, as envisioned by the tidal stirring scenario
\citep{kazantzidis11}.
Recently, \citet{lokas14} described in detail the formation and evolution of such a bar in a dwarf orbiting
in the gravitational field of the Milky Way.

An important aspect of the bar dynamics is the study of its orbital structure
to determine what types of orbits contribute to the density distribution of a bar.
It is believed that x$_1$ periodic orbits \citep[in the notation of][]{contopoulos80},
which are parallel to the bar, form the backbone of the bar \citep[for a review see][]{athanassoula13}.
This conjecture is supported by the classification of orbits in terms of $(\Omega-\Omega_\mathrm{p})/\kappa$,
where $\Omega$ is the angular frequency, $\kappa$ is the radial frequency of the orbit
and $\Omega_\mathrm{p}$ is the bar pattern speed.
In the inner part of the bar particles with $(\Omega-\Omega_\mathrm{p})/\kappa=1/2$ dominate
\citep{athanassoula02}, which is interpreted as the prevalence of x$_1$ orbits.

\section{Methods}

We reanalyzed the simulation of \citet{lokas14} which described the interaction of a dwarf galaxy
with a Milky Way-like host.
The dwarf model consisted of an NFW \citep{nfw95} dark matter halo of mass $10^9$~M$_{\odot}$ and a
stellar disk of mass $2 \times 10^7$ M$_{\odot}$.
The host galaxy model was made of an NFW halo of mass $7.7 \times 10^{11}$ M$_{\odot}$ and a
stellar disk of mass $3.4 \times 10^{10}$ M$_{\odot}$. The orbit of the dwarf was eccentric,
with the apocenter and pericenter of, respectively, $120$ and $25$ kpc. The dwarf was
initially placed at the apocenter and its evolution was followed using publicly available
$N$-body code \textsc{gadget2}  \citep{springel05}, using $4 \times 10^6$ particles in total.

As described in \citet{lokas14}, during the first pericenter passage a bar is formed in the disk of the dwarf. The length of the bar is about $2$ kpc.
At the following pericenters the bar is shortened and thickened.
For our analysis of the orbits we used only the period between the first and the second pericenter,
when the bar is the strongest. At this period, the bar rotates steadily for about $1.6$ Gyr so it is possible to use
a rotating reference frame, such that
the $x$, $y$ and $z$ axes are aligned with, respectively, the major, the intermediate
and the minor axis of the bar.
During this period of interest, we saved positions of the particles every $0.005$ Gyr to obtain
$320$ simulation outputs to probe the orbits.

We used the numerical analysis of fundamental frequencies \citep[see e.g.][]{valluri_merritt98}
which involves performing discrete Fourier transform of the time series of a given coordinate, e.g. $x(t)$.
We searched for the maximum of such a discrete spectrum and then refined the determination of the
location of the true peak of the Fourier spectrum. We designate the frequency of the peak as $\omega_x$.

\section{Results}

\begin{figure}
\centering
\includegraphics[width=0.8\textwidth]{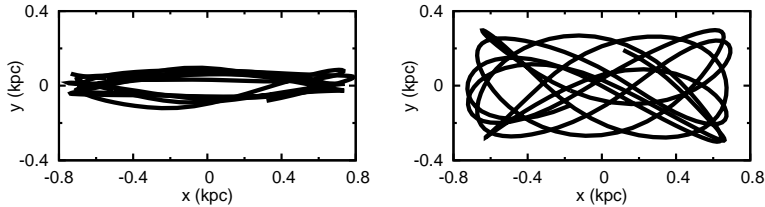}
\caption{\emph{Left:} Orbit of x$_1$ type, which has $\omega_y/\omega_x=1$.
\emph{Right:} Typical box orbit, with $\omega_y/\omega_x\approx 1.6$.
Both orbits have a mean radius of about $0.5$ kpc.}
\label{fig_orbits}
\end{figure}

In Figure \ref{fig_orbits} we show two typical shapes of orbits found in our simulation.
One of them (see the left panel of the Figure) is an x$_1$-type orbit, which is very elongated and circulates in
the same direction as the bar rotates. The second one (see the right panel) is a box orbit, also elongated with the bar,
but not as much as the x$_1$-type orbit.

\begin{figure}
\includegraphics[width=0.95\textwidth]{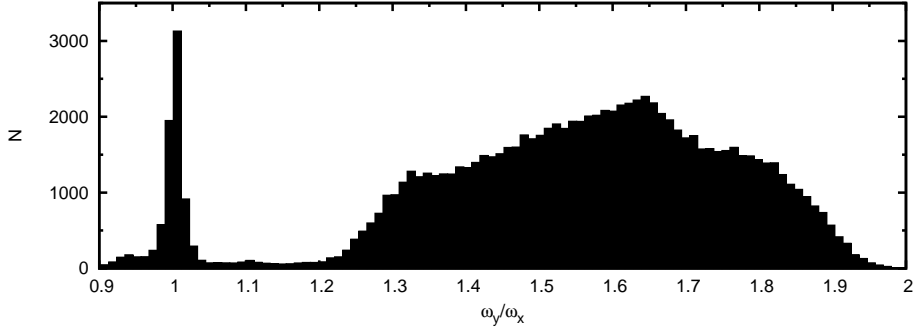}
\caption{Histogram of frequency ratios $\omega_y/\omega_x$ for $10^5$ orbits with the mean radius of $0.2$ -- $0.8$ kpc.}
\label{fig_freq_hist}
\end{figure}

We calculated the $\omega_x$ and $\omega_y$ frequencies for $10^5$ orbits with the mean distance
from the center of the dwarf in a range of $0.2$ -- $0.8$ kpc.
The lower boundary was set by the resolution of the simulation, whereas the upper one by the
lowest frequency we were able to resolve. The x$_1$-type orbits, which have mean distance from the center of $0.8$ kpc, extend along the bar up to $x\approx1.2$ kpc, thus our analysis refers to the inner half of the bar.
Figure \ref{fig_freq_hist} presents a histogram of frequency ratios $\omega_y/\omega_x$ for the analysed set of orbits.

The orbits clustered around $\omega_y/\omega_x=1$ are x$_1$-type orbits, however they contribute less than $10\%$ to the total
number of orbits.
The orbits in the range $\omega_y/\omega_x=1.2$--$1.9$ have boxy shapes. The ones around
$\omega_y/\omega_x\approx 1.3$ are almost square-like. Then, the larger $\omega_y/\omega_x$,
the more elongated is the orbit. The ones at $\omega_y/\omega_x\approx 1.9$ are almost as elongated
as the x$_1$ orbits.

\section{Summary}

We followed the orbits in a tidally induced bar in a simulated dwarf galaxy orbiting the Milky Way.
We computed fundamental frequencies of the particle orbits in the bar.
Only about $10\%$ of the orbits belong to x$_1$ type, whereas the majority of the orbits have boxy shapes,
with varying degree of elongation. Our results are thus in agreement with the
recent findings of \citet{patsis_katsanikas14},
who emphasized the importance of box orbits in bars.

\acknowledgements{This work was partially supported by the Polish National Science Centre under grant
2013/10/A/ST9/00023.}

\bibliographystyle{ptapap}
\bibliography{bibliography2}

\end{document}